\documentclass[12pt]{article}
\usepackage{epsfig,amsmath,amssymb,mathrsfs}
\usepackage[utf8]{inputenc}
\usepackage[colorlinks=true,citecolor=blue,,linktocpage=true,linkcolor=blue,urlcolor=black]{hyperref}
\usepackage{mathtools,slashed}
\usepackage{array}

\usepackage[force]{feynmp-auto}
\usepackage{caption}
\usepackage{subcaption}

  {\end{list}}%

\makeatletter
\renewcommand{\section}{\setcounter{equation}{0}\@startsection
 {section}%
 {1}%
 {0pt}%
 {-1\baselineskip}%
 {0.4\baselineskip}%
 {\bfseries\large}}%
\renewcommand{\subsection}{\@startsection
 {subsection}%
 {2}%
 {0pt}%
 {-0.75\baselineskip}%
 {0.2\baselineskip}%
 {\bfseries}}%
\renewcommand{\subsubsection}{\@startsection
 {subsubsection}%
 {3}%
 {0pt}%
 {-0.5\baselineskip}%
 {0.1\baselineskip}%
 {\sc}}%
\makeatother

\DeclareMathAlphabet{\mathpzc}{OT1}{pzc}{m}{it}

\def\be{\begin{equation}}
\def\ee{\end{equation}}

\def\g5{\gamma_{5}}

\def\id3k{\int\!\! \dfrac{d^3\!\vec{k}}{(2\pi)^3 2E(\vec{k})}}

\def\idx{\int\!\! d^4\!x}

\def\idx{\int\!\! d^{4}\!x}

\newcommand{\bea}{\begin{eqnarray}}
\newcommand{\eea}{\end{eqnarray}}
\newcommand{\beann}{\begin{eqnarray*}}
\newcommand{\eeann}{\end{eqnarray*}}
\newcommand{\ba}{\begin{array}}
\newcommand{\ea}{\end{array}}

 \def\g {\gamma}

\newcommand{\email}[1]{\href{mailto:#1}{\tt #1}}

\begin{document}

\rightline{\scriptsize{FTI/UCM 15-2020}}
\vglue 50pt

\begin{center}

{\LARGE \bf  A note on unimodular $N=1, d=4$ AdS supergravity}\\
\vskip 1.0true cm
{\Large Jesus Anero$^{\dagger}$, Carmelo P. Martin$^{\ddag}$ } {\large and} {\Large Raquel Santos-Garcia$^{\dagger}$}
\\
\vskip .7cm
{
	$^{\dagger}$Departamento de F\'isica Te\'orica and Instituto de F\'{\i}sica Te\'orica (IFT-UAM/CSIC),\\
	Universidad Aut\'onoma de Madrid, Cantoblanco, 28049, Madrid, Spain\\
	\vskip .1cm
	$^{\ddag}${Universidad Complutense de Madrid (UCM), Departamento de Física Teórica and IPARCOS, Facultad de Ciencias Físicas, 28040 Madrid, Spain}
	
	\vskip .5cm
	\begin{minipage}[l]{.9\textwidth}
		\begin{center}
			\textit{E-mail:}
			\email{jesusanero@gmail.com},
			\email{carmelop@fis.ucm.es}, \email{raquel.santosg@uam.es}
			
		\end{center}
	\end{minipage}
}
\end{center}
\thispagestyle{empty}

\begin{abstract}
We put forward a unimodular $N=1, d=4$ anti-de Sitter  supergravity theory off shell. This theory,  where  the Cosmological Constant does not couple to gravity, has a unique maximally supersymmetric classical vacuum which is Anti-de Sitter spacetime with radius given by the equation of motion of the auxiliary scalar field, ie, $S=\frac{3}{\kappa L}$. However, we see that the non-supersymmetric classical vacua of the unimodular theory are Minkowski and de Sitter spacetimes as well as anti-de Sitter spacetime with radius $l\neq L$.

\end{abstract}

\vfill
\clearpage

\section{Introduction}

The  $N=1, d=4$ supergravity theory  whose maximally supersymmetric classical vacuum is Anti-de Sitter (AdS) spacetime was introduced in reference \cite{Townsend:1977qa}. Today, supergravity on Anti-de Sitter spacetimes modulo some compact space play an important role in the gauge/gravity duality \cite{Ammon:2015wua, Nastase:2015wjb} program.

Unimodular gravity is a geometric theory of gravity which furnishes a Wilsonian solution to the problem of the huge discrepancy that exists between the experimental value of the Cosmological Constant and the theoretical value of the  latter as obtained within the quantum field theory framework. Indeed, in unimodular gravity the vacuum energy does not gravitate --see references \cite{vanderBij:1981ym} to \cite{deBrito:2019umw}, for further information.

A chief feature of unimodular gravity is that its gauge group is not the group of diffeomorphisms but a subgroup of it, namely: the group of transverse diffeomorphisms. Currently, gauge symmetries are viewed not as fundamental concepts but as  redundancies introduced so that locality is manifest in the Lagrangian formalism. These redundancies give rise to unphysical degrees of freedom and one has to show that they do not contribute to the observables of the theory. From this point of view, unimodular gravity has fewer redundancies than General Relativity and, hence, one has to deal with fewer unphysical degrees of freedom.

It has been shown in reference \cite{Anero:2019ldx} that there is a unimodular counterpart of $N=1, d=4$ Poincar\'e supergravity such that the supergravity algebra closes --off-shell-- on transverse diffeormorphisms, Lorentz transformations and unimodular supergravity transformations. As seen in \cite{Anero:2019ldx}, AdS is a non-supersymmetric vacua of the unimodular supergravity theory introduced
in \cite{Anero:2019ldx}. This is, of course, at odds with the situation that one finds in the standard --ie, non-unimodular-- $N=1, d=4$ Poincar\'e supergravity theory. It is thus interesting to see whether a unimodular  $N=1, d=4$ supergravity theory can be formulated so that Anti-de Sitter spacetime be a maximally supersymmetric vacuum of that theory.

The purpose of this note is to show that there is a unimodular  supergravity theory that  is, clearly, the unimodular counterpart of the $N=1, d=4$ AdS supergravity theory introduced in \cite{Townsend:1977qa}. Our starting point will be the unimodular gravity theory in \cite{Anero:2019ldx} --to keep the number of fields the same as in its standard counterpart, although there exist alternative formulations of unimodular $N=1, d=4$ supergravity --see references \cite{Nishino:2001gd, Nagy:2019ywi}, which involve the introduction of additional fields. In \cite{Nagy:2019ywi} de Sitter spacetime occurs as a supersymmetry breaking vacuum. The coupling of $3/2$-spin fermions to unimodular gravity has also been discussed in reference \cite{Blas:2008ce} --see reference \cite{Bonder:2018mfz} for Dirac fermions.

\section{Off-shell unimodular $N=1, d=4$ AdS supergravity}

Let $e^{a}_{\mu}$ and $\psi_\mu$ be the graviton vierbein and the gravitino field, respectively, which satisfy the following constraints
\begin{equation}
e\equiv\det\,e^a_\mu = 1,\quad \gamma^\mu\psi_\mu=0.
\label{uniconstraints}
\end{equation}
We define the action, $S_{UAdSSUGRA}$, of the off-shell unimodular $N=1, d=4$ AdS supergravity to be
\begin{equation}
S_{UAdSSUGRA}\,=\,S_{USG}\,+\, S_{L},
\label{Theaction}
\end{equation}
where
\begin{equation}
\begin{array}{l}
{S_{USG}=-\frac{1}{2\kappa^2}\!\!\idx\, e^{\mu}_a e^{\nu}_b R^{ab}_{\mu\nu}[\omega(e^c_\rho,\psi_\rho)]-\frac{i}{2}\!\!\idx\,\overline{\psi}_\mu\gamma^{\mu\nu\rho}D_{\nu}[\omega(e^a_\rho,\psi_\rho)]\psi_{\rho}-\frac{1}{3}\!\!\idx
\,\big[S^2+P^2+A^{a}A_{a}\big]}\\[4pt]
{S_{L}\,=\,\frac{2}{\kappa L}  \idx\, \big[S+ \frac{\kappa}{4} \bar{\psi}^\mu\psi_\mu\big].}
\end{array}
\label{Theactionbits}
\end{equation}
In the previous equations $S$, $P$ and $A_\mu$ are the auxiliary fields and $\gamma^{\mu_1\mu_2\mu_3}=\frac{1}{3!}\sum_{\pi}\,(-1)^{s_\pi}\,\gamma^{\mu_{\pi(1)}}\gamma^{\mu_{\pi(2)}}\gamma^{\mu_{\pi(3)}}$; $s_\pi$ being the signature of the permutation $\pi$. $S$ is a scalar field. $P$ is a pseudo-scalar field and $A_\mu$ is a pseudo-vector field.

Let us introduce the following infinitesimal transformations
\begin{equation}
\begin{array}{l}
{\delta_{\epsilon}e^a_{\mu}=-i\frac{\kappa}{2}\overline{\epsilon}\gamma^a\psi_{\mu},\quad \quad \gamma_{\mu}\equiv \gamma_a e^a_{\mu}}\\[4pt]
{\delta_{\epsilon}\psi_{\mu}=\frac{1}{\kappa}D_{\mu}[\omega^{ab}_\mu(e^c_{\sigma},\psi_{\sigma})]\epsilon
+\frac{i}{6}\gamma_{\mu}(S-i\gamma_5 P)\epsilon +\frac{i}{2}\gamma_5(\delta^{\nu}_{\mu}-\frac{1}{3}\gamma_{\mu}\gamma^{\nu})\epsilon  A_{\nu},}\\[4pt]
{\delta_{\epsilon}S=-\frac{1}{4}\overline{\epsilon}\gamma_{\mu}{\cal R}^{\mu},\quad\quad}\\[4pt]
{\delta_{\epsilon}P=\frac{i}{4}\overline{\epsilon}\gamma_5\gamma_{\mu}{\cal R}^{\mu}}\\[4pt]
{\delta_{\epsilon}A^a=\frac{3}{4}\overline{\epsilon}\gamma_5(\tilde{e}^a_{\nu}-\frac{1}{3}\gamma^a\gamma_{\nu})
{\cal R}^{\nu},}
\end{array}
\label{unisugratransfor}
\end{equation}
where
\begin{equation*}
\begin{array}{l}
{{\cal R}^\mu=\gamma^{\mu\nu\rho}{\cal D}_{\nu}\psi_{\rho},}\\[4pt]
{{\cal D}_{\mu}\psi_{\rho}=D_{\mu}[\omega^{ab}_{\nu}(e^c_{\sigma},\psi_{\sigma})]\psi_{\rho}
-i\frac{\kappa}{6}\gamma_{\rho}(S-i\gamma_5 P)\psi_{\mu} -i\frac{\kappa}{2}\gamma_5(\delta^{\lambda}_{\rho}-\frac{1}{3}\gamma_{\rho}\gamma^{\lambda})\psi_{\mu}A_{\lambda},}\\[4pt]
{D_{\mu}[\omega^{ab}_{\nu}(e^c_{\sigma},\psi_{\sigma})]=\partial_{\mu}+
\frac{1}{4}\omega(e^c_{\sigma},\psi_{\sigma})^{\phantom{\mu}ab}_{\mu}\gamma_{ab}.}
\end{array}
\end{equation*}
The symbol $\omega^{ab}_{\nu}(e^c_{\sigma},\psi_{\sigma})$ denotes the spin connection with torsion:
\begin{equation}
\begin{array}{l}
{\omega^{\phantom{\mu}ab}_{\mu}(e^c_\sigma,\psi_\sigma)=\omega^{\phantom{\mu}ab}_{\mu}(e^c_\sigma)+K^{\phantom{\mu}ab}_{\mu}(e^c_\sigma,\psi_\sigma),}\\[4pt]
{K^{\phantom{\mu}ab}_{\mu}(e^c_\sigma,\psi_\sigma)=i\frac{\kappa^2}{4}\big(\overline{\psi}_{\mu}\gamma^b\psi^a-\overline{\psi}^a\gamma_{\mu}\psi^b
+\overline{\psi}^b\gamma^a\psi_\mu\big);}
\end{array}
\label{spinconnection}
\end{equation}
$\omega^{\phantom{\mu}ab}_{\mu}(e^c_\sigma)$ being the Levi-Civita spin connection for the vierbein $e^a_{\mu}$.

It has been shown in reference \cite{Anero:2019ldx} that the  transformations in (\ref{unisugratransfor}) preserve the constraints in  (\ref{uniconstraints}) provided the infinitesimal parameter $\epsilon$ satisfies the following equation:
\begin{equation}
\gamma^{\mu}D_{\mu}[\omega(e^a_{\sigma},\psi_{\sigma})]\epsilon+i\frac{\kappa^2}{2} (\overline{\epsilon}\gamma^b\psi_a)\gamma^a\psi_b
+i\frac{2\kappa}{3}(S-i\gamma_5 P)\epsilon +i\frac{\kappa}{6}\gamma_5\gamma^{\nu}\epsilon A_{\nu}=0.
\label{preservingRSGdevelop}
\end{equation}
From now on we shall assume that $\epsilon$ in (\ref{unisugratransfor}) satisfies (\ref{preservingRSGdevelop}).

Let us show that $S_{UAdSSUGRA}$ in \eqref{Theaction} is invariant under the transformations in (\ref{unisugratransfor}). We have to show that the variation, $\delta_{\epsilon}S_{L}$ of $S_{L}$ under these transformations vanishes, for it has already been shown in  reference \cite{Anero:2019ldx} that $S$ in (\ref{Theactionbits}) is invariant under the transformations in question.

In the sequel we shall take advantage of the identity
\begin{equation}
\gamma^{\mu\nu}=\frac{i}{2}\,\epsilon^{\mu\nu\rho\sigma}\gamma_{\rho\sigma}\gamma_5,
\end{equation}
where $\epsilon^{0123}=1$ and $\gamma^{\mu\nu}=\frac{1}{2}[\gamma^\mu,\gamma^\nu]$.

It is not difficult to show that
\begin{equation}
\delta_{\epsilon}\idx\, S = -\frac{i}{4}\idx\, \epsilon^{\mu\nu\rho\sigma}\bar{\epsilon}\gamma_5\gamma_{\mu\nu}D_{\rho}\psi_{\sigma}-i\frac{\kappa}{4}\idx\,\bar{\epsilon}\gamma_5\psi^\lambda A_\lambda.
\label{Svar}
\end{equation}
Next, taking into account that $\delta_{\epsilon}(\gamma^\mu\psi_\mu)=0$, one gets
\begin{equation}
\begin{array}{l}
{\delta_{\epsilon}\idx\,\frac{\kappa}{4}\bar{\psi}^\mu\psi_\mu=-\frac{\kappa}{4}\,\delta_{\epsilon}\idx\,\bar{\psi}_\mu\gamma^{\mu\nu}\psi_\nu=}\\[4pt]
{-\frac{i}{4}\idx\,\epsilon^{\mu\nu\rho\sigma} (D_\mu\bar{\epsilon})\gamma_5\gamma_{\rho\sigma}\psi_\nu-\frac{\kappa^2}{8}\idx\,\epsilon^{\mu\nu\rho\sigma}(\bar{\epsilon}
\gamma^\lambda\psi_\rho)(\bar{\psi}_\mu\gamma_5\gamma_{\lambda\sigma}\psi_\nu)
+i\frac{\kappa}{4}\idx\,\bar{\epsilon}\gamma_5\psi^\lambda A_\lambda.}
\end{array}
\label{massvar}
\end{equation}

From (\ref{Theactionbits}), (\ref{Svar}) and (\ref{massvar}), one concludes that
\begin{equation*}
\delta_{\epsilon}S_{L}=\frac{2}{\kappa L}\big\{\frac{i}{4}\idx\,\epsilon^{\mu\nu\rho\sigma}\bar{\epsilon}\gamma_5\gamma_{cd}(D_\mu e^c_\rho-D_\rho e^c_\mu ) e^d_\sigma)\psi_\nu-\frac{\kappa^2}{8}\idx\,\epsilon^{\mu\nu\rho\sigma}(\bar{\epsilon}
\gamma^\lambda\psi_\rho)(\bar{\psi}_\mu\gamma_5\gamma_{\lambda\sigma}\psi_\nu)\big\}.
\end{equation*}
The previous equation boils down to
\begin{equation}
\delta_{\epsilon}S_{L}=-\frac{1}{4\kappa L}\idx\,\big[\epsilon^{\mu\nu\rho\sigma}(\bar{\epsilon}\gamma_5\psi_\sigma)(\bar{\psi}_\mu\gamma_\nu\psi_\rho)+
\epsilon^{\mu\nu\rho\sigma}(\bar{\epsilon}
\gamma^\lambda\psi_\rho)(\bar{\psi}_\mu\gamma_5\gamma_{\lambda\sigma}\psi_\nu)\big],
\label{almostzero}
\end{equation}
after using
\begin{equation*}
D_\mu e^c_\rho-D_\rho e^c_\mu= -i\frac{\kappa^2}{2}\bar{\psi}_\mu\gamma^c\psi_\rho.
\end{equation*}
Finally, applying a Fierz rearrangement to the second summand in (\ref{almostzero}), one gets that
\begin{equation*}
\delta_{\epsilon}S_{L}=0,
\end{equation*}
upon some Dirac algebra. We thus have shown that the transformations in (\ref{unisugratransfor}) leave the unimodular $N=1, d=4$ AdS action in (\ref{Theaction}) invariant provided the infinitesimal parameter
$\epsilon$ is a solution to equation (\ref{preservingRSGdevelop}).

It can be shown --the reader is referred to reference \cite{Anero:2019ldx} for the proof-- that if ${\cal F}$ denotes generically the fields in the transformations in (\ref{unisugratransfor}), and $\epsilon_1$ and
$\epsilon_2$ are infinitesimal parameters satisfying (\ref{preservingRSGdevelop}), then
\begin{equation}
[\delta_{\epsilon_1},\delta_{\epsilon_2}]{\cal F}=\delta^{(Diff)}_{\xi}{\cal F}+\delta^{(Lorentz)}_{\Lambda}{\cal F}+\delta_{\Sigma}{\cal F}.
\label{closedalgebra}
\end{equation}
Here $\delta^{(Diff)}_{\xi}$ is a transverse diffeomorphism with parameters $\xi^\mu$, $\delta^{(Lorentz)}_{\Lambda}$ denotes a Lorentz transformation with parameters $\Lambda^a_{\phantom{a}b}$ and $\delta_{\Sigma}$ is given by the  supergravity transformations in (\ref{unisugratransfor}) with parameter $\Sigma$ instead of $\epsilon$. $\xi^\mu$, $\Lambda^a_{\phantom{a}b}$ and $\Sigma$ are given by the
following equations
\begin{equation}
\begin{array}{l}
{\xi^\mu=\frac{i}{2}\overline{\epsilon}_1\gamma^\mu\epsilon_2,}\\[4pt]
{\Lambda^a_{\phantom{a}b}=\xi^\rho\omega_{\rho\phantom{a}b}^{\phantom{\rho}a}+\frac{\kappa}{6}\overline{\epsilon}_2\gamma^a_{\phantom{a}b}(S-i\gamma_5 P)\epsilon_1
-\frac{\kappa}{12}\overline{\epsilon}_2\{\gamma^a_{\phantom{a}b},\gamma^c\}\gamma_5\epsilon_1 A_c,}\\[4pt]
{\Sigma=\delta_{\epsilon_1}\epsilon_2-\delta_{\epsilon_2}\epsilon_1-\kappa \xi^{\rho}\psi_{\rho}.}
\end{array}
\label{parameters}
\end{equation}
$\xi^\mu$ is such that $\partial_\mu \xi^\mu=0$ and $\Sigma$ satisfies (\ref{unisugratransfor}). It is thus clear that the transformations in (\ref{unisugratransfor}) can be rightly called off-shell unimodular $N=1, d =4$ supergravity transformations.

\newpage

\section{Unimodular $N=1, d=4$ AdS supergravity and its classical solutions}

The equation of motion of the auxiliary fields $S$, $P$ and $A_a$  in the action in (\ref{Theaction}) read
\begin{equation}
S=\frac{3}{\kappa L},\quad P=0\quad\text{and}\quad A_a=0,
\label{auxeqom}
\end{equation}
respectively.
Let us remove the auxiliary fields $S$, $P$ and $A_a$ from the action in (\ref{Theaction}) by imposing their equations of motion. Thus, we obtain the following action
\begin{equation}
S^{(ON)}=-\frac{1}{2\kappa^2}\!\!\idx\, \big [ R\big[\omega(e^c_\rho,\psi_\rho)]+2\Lambda\big]-\frac{i}{2}\!\!\idx\,\overline{\psi}_\mu\gamma^{\mu\nu\rho}D_{\nu}[\omega(e^a_\rho,\psi_\rho)]\psi_{\rho}
+\frac{1}{2 L}  \idx\, \bar{\psi}^\mu\psi_\mu,
\label{onshellaction}
\end{equation}
where the $\Lambda=-3/L^2$ and $e=1$ and $\gamma^\mu\psi_\mu=0$. Notice that the Cosmological Constant does not couple to gravity as befits a unimodular theory of gravity.

The supergravity transformations that leave $S^{(ON)}$ invariant are obtained by imposing the equation of motion in (\ref{auxeqom}) on the transformations in (\ref{unisugratransfor}). Thus one obtains
\begin{equation}
\begin{array}{l}
{\delta_{\epsilon}e^a_{\mu}=-i\frac{\kappa}{2}\overline{\epsilon}\gamma^a\psi_{\mu},\quad \quad \gamma_{\mu}\equiv \gamma_a e^a_{\mu}}\\[4pt]
{\delta_{\epsilon}\psi_{\mu}=\frac{1}{\kappa}D_{\mu}[\omega^{ab}_\mu(e^c_{\sigma},\psi_{\sigma})]\epsilon
+\frac{i}{2\kappa L}\gamma_{\mu}\epsilon.}
\end{array}
\label{unisugraonshell}
\end{equation}
Of course, the infinitesimal parameter, $\epsilon$, in (\ref{unisugraonshell}) satisfies now the equation
\begin{equation}
\gamma^{\mu}D_{\mu}[\omega(e^a_{\sigma},\psi_{\sigma})]\epsilon
+i\frac{2}{L}\epsilon=-i\frac{\kappa^2}{2} (\overline{\epsilon}\gamma^b\psi_a)\gamma^a\psi_b.
\label{preservingonshell}
\end{equation}

The classical solutions --the classical vacua-- of the theory whose action is $S^{(ON)}$ in (\ref{onshellaction}) are those field configurations, $(e^a_\mu,\psi_\mu)$, for which  the action $S^{(ON)}$ is stationary and $\psi_\mu=0$. Hence, these field configurations $(e^a_\mu,\psi_\mu=0)$ are solutions to the unimodular gravity equations in vacuum
\begin{equation}
R_{\mu\nu}-\frac{1}{4} R g_{\mu\nu}=0.
\label{unieq}
\end{equation}
$R_{\mu\nu}$ and $R$ denote the Ricci tensor and the scalar curvature for the unimodular metric $g_{\mu\nu}= e^a_\mu e_{a\nu}$ , respectively. It is well known that the space of solutions to (\ref{unieq}) is the space of Lorentzian Einstein manifolds. Hence, Minkowski, de Sitter and Anti-de Sitter are classical vacua of the theory with action $S^{(ON)}$ in (\ref{onshellaction}). However, both Minkowski and de Sitter spacetimes break supersymmetry, whereas Anti-de Sitter with radius $L$ is maximally supersymmetric. Indeed, given a gravitational background, $e^a_\mu$, $\psi_\mu=0$ is invariant under a supergravity transformations with parameter $\epsilon$ if and only if
\begin{equation}
0=\kappa\delta_{\epsilon}\psi_{\mu}=D_{\mu}[\omega^{ab}_\mu(e^c_{\sigma},\psi_{\sigma})]\epsilon
+\frac{i}{2 L}\gamma_{\mu}\epsilon\quad \Longleftrightarrow\quad D_{\mu}[\omega^{ab}_\mu(e^c_{\sigma},\psi_{\sigma})]\epsilon+\frac{i}{2L}\gamma_\mu\epsilon=0.
\label{killingspinoreq}
\end{equation}
Hence, $\epsilon$ must be a Killing spinor of the gravitational background in question. Notice that when $\psi_\mu=0$, the Killing spinor equation in  (\ref{killingspinoreq}) implies equation (\ref{preservingonshell}), so there is no clash between them.

It is well known that for Minkowski and de Sitter spacetimes no $\epsilon$ exists that solves the killing equation in (\ref{killingspinoreq}). Hence, they are not supersymmetric vacua. On the other hand, if the gravitational background is Anti-de Sitter with radius $L$, there exists a maximal number of independent solutions to (\ref{killingspinoreq}). Notice that if the radius, l, of the anti-de Sitter spacetime is not $L$, we still have a solution to (\ref{unieq}) --ie, a classical vacuum, but, this vacuum breaks supersymmetry in a maximal way. Indeed, the integrability condition for the killing equation in (\ref{killingspinoreq})
reads
\begin{equation}
0=[D_\mu+\frac{i}{2L}\gamma_\mu,D_\nu+\frac{i}{2L}\gamma_\nu]\epsilon=\big[R^{ab}_{\mu\nu}-\frac{1}{L^2}(e^a_\mu e^b_\nu-e^a_\nu e^b_\mu)\big]\gamma_{ab}\epsilon,
\end{equation}
whereas we have $R^{ab}_{\mu\nu}=\frac{1}{L^2}(e^a_\mu e^b_\nu-e^a_\nu e^b_\mu)$ for a $d=4$ Anti-de Sitter spacetime with radius $L$.

\section{Conclusions and outlook}

We have shown that there exists a unimodular $N=1, d=4$ AdS supergravity where the supergravity transformations involve a length $L$. This theory has AdS with radius $L$ as its maximally supersymmetric classical vacuum. An yet, Minkowski spacetime, de Sitter space-time and anti-de Sitter with radius $l\neq L $ are classical vacua of the theory, which --spontaneously-- break supersymmetry in a maximal way. Further, as in non-supersymmetric unimodular gravity the vacuum energy does not gravitate.

It will be very interesting to  develop a superconformal approach to unimodular supergravity by adapting or generalizing the standard --see Chapter 16 of \cite{Freedman:2012zz}-- approach to supergravity. Thus one will be able to properly compare the family of theories --parametrized by $L$, the AdS radius-- presented here with the family of supergravity theories constructed in \cite{Bergshoeff:2015tra, Hasegawa:2015bza}. Notice that the family of supergravity theories in the unitary gauge and parametrized by $f$ and $m$ --see eq. (4.10) of \cite{Bergshoeff:2015tra}-- presented in \cite{Bergshoeff:2015tra, Hasegawa:2015bza} admits Minkowski and AdS as vacua, these vacua are always maximally supersymmetric; wehereas in the unimodular supergravity theory introduced in this paper, Minkowski spacetime and AdS with appropriate radius are vacua which break supersymmetry. It would appear that the family of unimodular supergravity theories we have put forward here is not altogether equivalent to the family of theories in \cite{Bergshoeff:2015tra, Hasegawa:2015bza}. However, as we have just said, to settle this issue properly will demand the construction of a superconformal approach to unimodular supergravity and this lies outside the scope of this paper.

\section{ Acknowledgements}

We are very much indebted to Prof. E. \'Alvarez for countless enlightening discussions on unimodular supergravity.  This work has been partially supported by the Spanish Ministerio de Ciencia, Innovaci\'on y  Universidades through grant PGC2018-095382-B-I00. This work has also received funding from the Spanish Research Agency (Agencia Estatal de Investigacion) through the grant IFT Centro de Excelencia Severo Ochoa SEV-2016-0597, and the European Union's Horizon 2020 research and innovation programme under the Marie Sklodowska-Curie grants agreement No 674896 and No 690575. We also have been partially supported by FPA2016-78645-P(Spain). RSG is supported by the Spanish FPU Grant No FPU16/01595.


\end{document}